# Accelerated molecular dynamics simulation of low-velocity frictional sliding

W K Kim[1] and M L Falk[1,2,3]

[1] Department of Materials Science and Engineering, Johns Hopkins University, Baltimore, MD 21218, USA

[2] Department of Mechanical Engineering, Johns Hopkins University, Baltimore, MD 21218, USA

[3] Department of Physics and Astronomy, Johns Hopkins University, Baltimore, MD 21218, USA

**Abstract**

Accelerated molecular dynamics (MD) simulations are implemented to model the sliding process of AFM experiments at speeds close to those found in experiment. In this study the hyperdynamics method, originally devised to extend MD time scales for non-driven systems, is applied to the frictional sliding system. This technique is combined with a parallel algorithm that simultaneously simulates the system over a range of slider positions. The new methodologies are tested using 2-dimensional and 3-dimensional Lennard-Jones AFM models. Direct comparison with the results from conventional MD shows close agreement validating the methods.

## 1. Introduction

While friction has presented many intriguing challenges since the onset of civilization [1], only in recent years have researchers been able to consider the origin of friction by directly investigating atomic-scale interactions. This change has been triggered mainly by the invention of Atomic Force Microscope (AFM) [2], which made it possible to measure the friction force acting on a single-asperity contact at the nano-meter scale. Moreover, the development of new hardware and computer simulation methodologies have played an important role in interpreting the origin of frictional forces and energy dissipation during sliding at this scale [3-8].

In an AFM experiment, only a small number of atoms at the end of a tip are in contact with the atoms in the substrate. The common physical picture of the process of frictional sliding is as follows. The atoms at the tip are initially equilibrated at a local free energy minimum, but as the slider moves and the cantilever deforms the initial minimum configuration becomes meta-stable. Due to thermal fluctuations, the atoms rearrange into a new more stable minimum releasing the elastic energy stored in the cantilever. This process: equilibration at a local minimum, escape from the minimum and establishment of a new local minimum, is repeated as the slider advances.

Molecular dynamics (MD) simulations have been used to model this process, but it has not been possible to perform the simulations at sliding velocities close to typical experimental values (nm/s ~ μm/s) due to MD's short time scale limited to sub-microseconds and velocities in excess of meters per second. In recent years, several novel methods to extend the MD time scale have been devised [9-13], but most of these are not suitable for driven systems, whose boundary conditions change with time. One notable exception is the extension of the parallel replica method for driven systems: this method has been applied to straining nano-tubes and stick-slip friction [14, 15]. However, the parallel replica method has only allowed a decrease in sliding speed by a factor corresponding to the number of processors, which is insufficient to reach experimental sliding speeds.

In this paper we present a novel method to extend MD time scale for driven systems using hyperdynamics [9, 10] and an alternative parallel algorithm. Like the parallel replica method for driven systems [14, 15], the fundamental assumption of our method is that the system boundary conditions change slowly so that the system remains at a near equilibrium state. We apply the method here to a sliding system with a slider moving at a constant velocity. Instead of simulating multiple independent replicas simultaneously, we parallelize slider positions and perform the simultaneous hyperdynamics simulations with the models at different slider positions.

## 2. Method

In this section we will briefly review transition state theory (TST) and the hyperdynamics method. The hyperdynamics method is based on the TST assumptions. Our methodology is described in Sec. 2.3.

### 2.1. Infrequent events and transition state theory

Most of the kinetic phenomena macroscopically observed in solid materials such as diffusion and creep are related to thermally activated changes in the configuration of the atoms comprising the material. Atoms make transitions from one meta-stable potential energy basin to another when they gain enough energy to overcome the energy barrier due to thermal fluctuations. In these dynamical systems, the information about the waiting times at each state and the transition mechanisms leading to other states as well as their relative probabilities is essential to understand the underlying physics of the phenomena.

In many cases, before hopping to other states, the system stays in the neighborhood of a potential energy minimum for a very long time compared to the typical atomic vibration period and a transition itself occurs in a relatively short time. This long waiting time hinders the utility of conventional MD methodologies. In these situations, the transition rate is an equilibrium property that is independent of specific trajectories of the system and the waiting time at each state has a Poisson distribution expressed as

$$p(t) = R \exp(-R t) , \tag{1}$$

where $R$ is a rate constant that characterizes the transition.

Transition state theory can provide an analytical expression for the transition rates in an infrequently hopping system so long as we can construct a proper dividing surface that the system crosses in transitions [16, 17]. Then, the transition rate at a state $A$ is given by

$$R_A^{TST} = \frac{1}{2} \frac{\int d\vec{r} \int d\vec{v} \, |v_n| \, \delta_S(\vec{r}) \exp[-\beta(V+K)]}{\int_A d\vec{r} \int d\vec{v} \exp[-\beta(V+K)]} , \tag{2}$$

where $\vec{r}$ is the 3$N$-dimensional position vector in the configuration space ($N$ is the total number of particles), $\vec{v}$ is the 3$N$-dimensional velocity vector, $v_n$ is the velocity normal to a dividing surface $S$, $\delta_S(\vec{r})$ is a Dirac-delta function located at the surface, $V$ is the potential energy, $K$ is the kinetic energy, and $\beta = 1/k_B T$; $k_B$ is the Boltzmann constant and $T$ is the temperature. In many cases the dividing surface is chosen as a hyper-plane passing through the saddle point between two given minima and normal to the eigenvector corresponding to the lowest eigenvalue of the Hessian at the saddle point.

*2.2. Hyperdynamics for non-driven system*

In this section we briefly review the hyperdynamics method and more detailed description is found in [9]. Based on the assumptions of TST, hyperdynamics uses a potential modified from a given potential to reduce the energy barriers. As long as the modified potential does not alter the original potential along the TST dividing surface, it can be proven that the relative probabilities to neighboring states are identical in both potentials [9]. Then, the ratio of the transition rates in these potentials at a given state $A$ is given by

$$\alpha = \frac{\left(R_A^{TST}\right)_b}{R_A^{TST}} = \frac{\bar{t}_A}{\left(\bar{t}_A\right)_b} = \frac{\int_A \exp[-\beta V] d\vec{r}}{\int_A \exp[-\beta V_b] d\vec{r}} = \left\langle e^{+\beta \Delta V} \right\rangle_b , \tag{3}$$

where $\alpha$ is the boost factor, $\bar{t}$ is an average waiting time that is the inverse of the transition rate and $V_b = V + \Delta V$, and $\langle \cdots \rangle_b$ is the ensemble average in the modified potential. The bias potential $\Delta V$ must satisfy the following condition.

$$\Delta V(\vec{r}) = \begin{cases} > 0, & \text{in } A \\ = 0, & \text{along } S \end{cases} . \tag{4}$$

The difficulty of performing a hyperdynamics simulation arises particularly from the subtlety of constructing a computationally efficient bias potential. Voter's original bias potentials used the lowest eigenvalue and the corresponding eigenvector of the Hessian matrix [9, 10], but calculating the eigenvalue and its derivative require significant computational overhead.

For this study we have devised new bias potentials. These bias potentials use local variables to approximate a TST dividing surface. If a local variable is bounded along the dividing surface, the variable can be used to construct a volume located inside the dividing surface. For example, the lowest eigenvalue of the Hessian matrix is negative at the dividing surface so that the configuration volume with positive lowest eigenvalues can be located inside the dividing surface. We considered a number of local variables that could be used in place of or in addition to the lowest eigenvalue of the Hessian. Possible choices are the potential energy slope or curvature along the direction vector connecting a configuration and the minimum of the potential energy basin, and the distance from this minimum. They are defined by

$$\sigma = \nabla V \cdot \vec{s} = \frac{dV}{ds} \, , \tag{5}$$

$$\kappa = \frac{d^2 V}{ds^2} \, , \tag{6}$$

and

$$d = \sqrt{\sum_{k=1}^{3N} (r_k - r_{O,k})^2} \, , \tag{7}$$

where $\vec{r}_O$ is the generalized 3N-dimensional position vector at the local minimum, $\vec{s} = (\vec{r} - \vec{r}_O)/d$, $r_k$ is a component of the position vector $\vec{r}$ and $r_{O,k}$ is a component of $\vec{r}_O$. In some specific systems, the slope in equation (5) and/or the curvature in equation (6) become smaller and the distance in equation (7) increases when the system approaches a dividing surface. In such cases it may be possible to determine critical values for them. A more detailed review regarding the methodology we use to construct bias potentials is in preparation [18].

*2.3. New acceleration method for driven system: Hyperdynamics with a parallel algorithm*

The original hyperdynamics method was developed assuming a time-invariant potential energy. In this study we extend the method for driven systems. Moreover, we incorporate this method into a parallel algorithm.

As boundary conditions in a driven system change with time, the system is not, strictly speaking, in equilibrium. However, as we will detail below, if the change rate is so slow that the system remains at a near-equilibrium state, then we can still apply the rate theory based on equilibrium assumptions and the instantaneous transition rate can be calculated using equation (2) with an instantaneous potential function.

In our AFM model the external parameter changing in time is the slider position $x_S$ and the potential energy is a function of the slider position as well as the atomic positions. By assuming that the sliding velocity is low enough, the slider position is updated by $\Delta x_S$ after a time period $\Delta\tau$ has elapsed instead of changing continuously as shown in figure 1. $\Delta\tau$ is determined by the sliding rate ($\Delta\tau = \Delta x_S / v_S$). For this process to be equivalent to the continuous sliding, $\Delta x_S$ should be small compared to the length scale that characterizes the surface corrugation. Moreover, $\Delta\tau$ must be longer than the thermal equilibration time-scale ($\Delta\tau \gg \tau_{eq}$).

We turn to the interpretation of the hyperdynamics simulations within the stepped sliding scheme. In the stepped sliding, if we use the original potential with the slider, frozen for $\Delta\tau$, we have to run the simulation for the same period $\Delta\tau$. Then, the probability for the transition for this time period is given by

$$1 - e^{-R\Delta\tau}, \tag{8}$$

where $R$ is the transition rate corresponding to the slider position. If we perform the same simulation with a biased potential, which has the boost factor of $\alpha$, then the transition rate increases ($R_b = \alpha \times R$). Note that we have the same probability for the shorter time period ($\Delta\tau_b = \Delta\tau / \alpha$) because $1 - e^{-R\Delta\tau} = 1 - e^{-(R\alpha)(\Delta\tau/\alpha)} = 1 - e^{-R_b \Delta\tau_b}$. Therefore, with the biased potential we can reduce the simulation time.

Assuming that the slider moves so slowly that the system is fully equilibrated at each slider position, the dynamics at different slider positions are uncorrelated. Thus, the transition probabilities at each slider position during stepped sliding are independent of each other, and we do not necessarily need to perform the simulations successively. Rather, we can perform the simulations with different slider positions in parallel as illustrated in figure 2. For example, if we use the conventional serial algorithm for the system shown in figure 2, we have to first perform a simulation at $x_S = 1$ (figure 2 (a)), and after it finishes, we perform another simulation at $x_S = 2$ (figure 2 (b)), etc. until we observe a transition. This corresponds to throwing dice and throwing again after knowing the first result. However, if these two events are independent of each other, we can throw both simultaneously. Thus, we can perform four simulations at $x_S = 1, 2, 3, 4$, simultaneously. If we have a transition at $x_S = 3$ and this is the latest slider position that experienced a transition in the time interval, then we ignore the result at $x_S = 4$ and redistribute the jobs starting from $x_S = 3$ and restart to perform the simulations. The speed-up obtained by this parallel distribution method is roughly proportional to the number of processors used for one system.

## 3. Application

*3.1. 2-Dimensional L-J system*

*3.1.1 Model*

We begin by investigating a simple 2-dimensional AFM model illustrated in figure 3. The substrate consists of 80 atoms marked in blue, and the tip contains 33 atoms marked in red. The tip and the substrate have a 2-dimensional crystalline structure corresponding to an FCC crystal in 3-dimension. The lattice parameters of the tip and the substrate are identical. Thus, the tip atoms contacting the substrate are located at the lattice sites of the substrate as shown in figure 3.

The atoms on the bottom layer of the substrate are fixed to prevent a rigid-body translation in the vertical direction, and the system is subject to the periodic boundary condition in the horizontal direction. The relative motions of the atoms on the top layer of the tip are constrained, but they can move like a rigid body. These top atoms are pulled by a spring and pushed downward by the applied normal force as shown in figure 3. All the quantities are expressed in length units of $\sigma$, the energy units of $\varepsilon$, and the mass units of $m$. Time is measured in the time units of $\tau = \sqrt{m\sigma^2/\varepsilon}$. Hereafter the units are omitted unless there is ambiguity. We used a spring stiffness of $k = 5$ and an applied normal force of $F_N = 5$.

The interactions of the atoms are modeled by the Lennard-Jones potential,

$$V(r) = 4\varepsilon_{ab}\left[\left(\frac{\sigma_{ab}}{r}\right)^{12} - \left(\frac{\sigma_{ab}}{r}\right)^{6}\right], \tag{9}$$

where $\varepsilon_{ab}$ is the bond energy between the atom of the type $a$ and the atom of the type $b$, $\sigma_{ab}$ is the characteristic length parameter, and $r$ is the distance between the two atoms. We used the following parameters.

$$\sigma_{ss} = \sigma_{tt} = \sigma_{ts} = 1.0, \varepsilon_{ss} = \varepsilon_{tt} = 1.0, \varepsilon_{ts} = 0.5 \quad (s: \text{substrate}, t: \text{tip})$$

Note that we used a smaller value of the bond energy for the interaction between the tip and the substrate to guarantee that the slip always occurs at the interface rather than inside the tip.

To validate our methodologies we tested four different methods; (1) continuous sliding on a single processor, (2) stepped sliding on a single processor, (3) stepped sliding using the parallel method, and (4) stepped sliding using a biased potential (hyperdynamics) with the parallel method. In case of stepped sliding, we update the slider position by $\Delta x_S = 0.01$ and the time period $\Delta \tau$ (= $\Delta x_S / v_S$) during which the slider position is fixed increases as the sliding velocity decreases. For the hyperdynamics simulations, we used a bias potential using a local slope defined in equation (5).

We varied the sliding velocity by 5 orders of magnitude ranging from $v_S = 10^{-4}$ to $10^{-8}$ and three different temperatures ($T = 0.1, 0.01, 0.001 \ \varepsilon/k_B$) have been simulated using the Nose-Hoover chain method [19]. The equations of motion are solved using a modified velocity-Verlet algorithm [20].

*3.1.2. Results*

The graphs shown in figure 4 through figure 6 are obtained from a simulation with $v_S = 10^{-6}$ and $T = 0.01$ and illustrate typical frictional behaviors of the model.

Figure 4 shows the tip position $x_T$, measured at the top layer, as a function of the slider position $x_S$, and the slider position is plotted together with the tip position. As in the Tomlinson model, apparent stick-slip motion is observed. The tip position increases linearly during the stick-phase and jumps at several discrete points corresponding to slip events. The average distance of these points corresponds to the lattice parameter of the substrate. Note that the tip position shown in the figure is the averaged quantity over a time period (otherwise the curve is very noisy due to thermal fluctuations).

The lateral force $F_R$ is measured by the deformation of the spring, as in an AFM experiment, expressed as

$$F_R = k(x_S - x_T) \ . \tag{10}$$

Figure 5 (a) and 5 (b) show the lateral force at the first slip as a function of the slider position and the tip position respectively. As expected from figure 4, where the tip position is linearly proportional to the slider position ($x_T \approx k_C x_S$), the lateral force exhibits a linear dependence on both the slider position and the tip position. However, the straight line extended from the initially linear portion of the curves illustrates that the lateral force deviates from the linear dependence near the transition.

The potential energy, which is a function of the tip position as well as the atom positions, $V(\vec{r}_1, \cdots, \vec{r}_N; \vec{r}_T)$, is shown in figure 6. Figure 6 (a) shows the potential energy as a function of the slider position, and figure 6 (b) shows the potential energy as a function of the tip position. The increase in the potential energy is due to the elastic deformation of the tip and can be fit to a quadratic function. We expect that the time averaged potential energy $\overline{V}$ has the following relation with the slider position.

$$\overline{V} \sim \frac{1}{2} k_1 \, x_S^2 \ , \tag{11}$$

where $k_1$ is a constant. The linear and quadratic increases in the tip position and in the potential energy respectively imply a linear decrease in the energy barrier as the slider advances. However, Catastrophe theory predicts [21, 22] that the energy barrier decreases as a function of $(x_S^* - x_S)^{3/2}$ near the transition, where $x_S^*$ is the slider position where the energy barrier completely vanishes. Thus, as the slider approaches near transition we could expect the deviations. A very small deviation is apparent in the tip position as shown in figure 5 and no deviation in the potential energy is evident in figure 6.

Figure 7 shows the dependence of the lateral force on the sliding velocity. At a temperature of 0.01 the results from five different sliding velocities ($10^{-4}$, $10^{-5}$, $10^{-6}$, $10^{-7}$, $10^{-8}$) are shown. It is apparent that as the sliding velocity decreases the tip makes a transition at an earlier slider position, which is consistent with the prediction of the modified Tomlinson model [23]. The temperature dependence is shown in figure 8. In this figure we can observe that the transition occurs at much earlier slider position at higher temperature, and the effective stiffness $k_{eff}$, the slope of the lateral force vs. slider position curve, slightly reduces as temperature increases due to softening of the tip and contact stiffness.

Finally, we compare the results from various methods. Figure 9 summarizes the simulation results at various sliding velocities and various temperatures obtained from the four different methods. At each velocity and temperature, we prepared 10 samples for serial simulations and 5 samples for parallel simulations. Each sample has different initial conditions. The lateral forces in this graph are measured at the transition points and averaged over eight different transition points and over different samples.

The continuous sliding and the stepped sliding (using the original potential on a single processor) are tested at the velocities of $10^{-4}$, $10^{-5}$, $10^{-6}$ and at temperatures of 0.001, 0.01, 0.1. All the data overlap and agree within the range of the standard deviation shown as the error bar in figure 9. Most data ranges within the standard error.

At the sliding velocity of $10^{-6}$ and the temperature of 0.01, all four methods are tested and all the measured lateral forces agree. The simulations on a single processor could not have been performed at the sliding velocities lower than $10^{-6}$ due to excessive running time on a standard workstation, but all the data from the parallel simulations at $10^{-7}$ and $10^{-8}$ using either the original potential or the biased potential fall close to the trend line extended from the data obtained from continuous sliding.

The lateral forces show the expected logarithmic dependence on the sliding velocity, and no plateaus are found in any velocity range. The slope of the lateral force vs. ln $v_S$ curve increases as temperature increases.

*3.1.3. Discussion*

With the conventional method, we were not able to perform simulations at velocities lower than $10^{-6}$ because of the extended running time on a standard workstation. With the velocities above this limit, the simulation results from the stepped sliding agree with the results of the continuous sliding. Thus, the basis for the other methods (the parallel method and the hyperdynamics methods) is well verified. Using the parallel method makes it possible to lower the sliding velocity by one order of magnitude, and with the hyperdynamics methodology we can lower the sliding velocity further.

In this 2-dimensional sliding system, we have found that the relative population density of the unboosted region in the phase space is altered as the slider position changes. When the slider position is far from the transition point, the pre-simulation using the original potential to calculate the boost factor does not sample any points in the unboosted region. Thus, the maximum boost factor, which is the inverse of the relative population density of the unboosted region, will be very large [18]. However, as the slider approaches the transition point, some unboosted points are sampled and the maximum achievable boost factor reduces. Since as the sliding velocity decreases the transition occurs at earlier slider positions where the maximum boost factor is larger, we expect that we can reduce the sliding velocity even below $10^{-8}$.

Although the lateral force shows a logarithmic dependence on the sliding velocity, this is expected due to the simplicity of the current model. Since the tip maintains its crystalline structure after transitions and no defects arise inside the tip due to much weaker interaction between the tip and the substrate, the only possible transition mechanisms are backward and forward hopping, which have the same energy barriers. As the slider advances, the forward hopping (in the sliding direction) becomes more favorable than the backward hopping. Moreover, the relative configurations of the system before and after transition do not change. However, in more realistic situations, the tip may lose atoms during sliding and its interface configuration may be altered during transitions or different pathways may be traversed at high and low temperature.

## 3.2. 3-Dimensional L-J system

### 3.2.1. Model

We now proceed to a 3-dimenstional system modeling an AFM tip and a substrate illustrated in figure 10. The tip has 183 atoms shown in red and the substrate consists of 1800 atoms shown in blue. The substrate has FCC crystalline structure, and the tip is created by carving an FCC crystal with the same lattice parameter as the substrate into a conical shape with flat ends. The tip and the substrate are joined in the [001] direction, and as shown on the right side of figure 10, nine atoms on the bottom of the tip are in contact with the substrate. Because the tip and the substrate have the same lattice parameter and are aligned in the same orientation the tip atoms are in registry with the substrate.

The sliding simulation is realized in the same way as the 2-dimensional model. A spring ($k = 10$) is linked to the top layer of the tip and the bottom layer of the substrate is fixed. A normal force ($F_N = 5$) is applied to the top of the tip.

The interaction between substrate atoms and the interaction between a substrate atom and a tip atom are modeled by the Lennard-Jones (L-J) potential, and the following parameters are used.

$$\sigma_{ss} = 1.0 \quad , \quad \varepsilon_{ss} = 1.0$$
$$\sigma_{st} = 1.0 \quad , \quad \varepsilon_{st} = 0.2$$

(s: substrate, t: tip)

For the interaction between tip atoms, we used a harmonic potential, which does not allow any bond breaking to maintain the shape of the tip and prevent wear during sliding.

$$V(r) = \frac{1}{2} k (r - r_O)^2 , \qquad (12)$$

where $k$ is the stiffness (= 57.2), and $r_O$ is the equilibrium bond length (= 1.12). The stiffness and the equilibrium length are chosen to be identical to the values of the L-J potential with $\sigma_{tt} = 1$, $\varepsilon_{tt} = 1$ at the equilibrium position.

We tested the same four methods as in 2-D simulations, and for the hyperdynamics simulations we used a bias potential using a local slope $\sigma$ defined in equation (5) and the lowest eigenvalue $\varepsilon$ of the Hessian Matrix.

$$\Delta V(\sigma, \varepsilon) = \Delta V_1(\sigma) + \Delta V_2(\varepsilon) - \frac{1}{\Delta V_{max}} \Delta V_1(\sigma) \times \Delta V_2(\varepsilon) , \qquad (13)$$

where

$$\Delta V_1(\sigma) = \begin{cases} \Delta V_{max} & \sigma \geq \sigma_U \\ \Delta V_{max} \left[ 1 - \left( \frac{\sigma - \sigma_U}{\sigma_L - \sigma_U} \right)^m \right]^2 & \sigma_L < \sigma < \sigma_U \\ 0 & \sigma \leq \sigma_L \end{cases} \qquad (14)$$

and

$$\Delta V_2(\varepsilon) = \begin{cases} \Delta V_{max} & \varepsilon \geq \varepsilon_U \\ \Delta V_{max} \left[ 1 - \left( \frac{\varepsilon - \varepsilon_U}{\varepsilon_L - \varepsilon_U} \right)^m \right]^2 & \varepsilon_L < \varepsilon < \varepsilon_U \\ 0 & \varepsilon \leq \varepsilon_L \end{cases} , \qquad (15)$$

where $\sigma_L$ and $\sigma_U$ are the lower and upper critical values for the slope respectively, $\varepsilon_L$ and $\varepsilon_U$ are the lower and upper critical values for the eigenvalue respectively, and $m$ is an integer. Note that $\Delta V(\sigma, \varepsilon)$ ranges from 0 to $\Delta V_{max}$.

We performed simulations at four different sliding velocities ($10^{-4}$, $10^{-5}$, $10^{-6}$, $10^{-7}$) and at temperatures of 0.001, 0.01, 0.1. As in the 2-dimensional model, we used the Nose-Hoover chain method to control temperature [19] and a modified velocity-Verlet algorithm to numerically solve the equations of motion [20].

*3.2.2. Results*

We plotted the lateral force and the potential energy as functions of both the slider position and the tip position in figure 11 and figure 12, where the data are obtained from the simulations with $v_S = 10^{-5}$ and $T = 0.01$. As in the 2-dimensional case, the lateral forces show the linear dependence, but deviates from the straight lines near transition points (figure 11). The potential energy changes like a quadratic function of the slider position at earlier slider positions, but shows deviation from the quadratic fits unlike the 2-D models and very steep changes near transitions (figure 12).

Figure 13 summarizes the simulation results. The lateral forces shown in the figure are the averages of the peak values at each transition over the samples and the peaks. At sliding velocities of $10^{-4}$ and $10^{-5}$, the lateral forces measured from the continuous sliding show close agreement with the forces from the stepped sliding. Thus, the fundamental assumption of our methodologies is verified with this 3-dimensional model. However, although the number of atoms in this model (1,983) is not large, the simulations on a single processor using the conventional method at lower sliding velocities ($< 10^{-5}$) are prohibitive, requiring more than one month on a standard workstation.

By the parallel method using 50 processors, we were able to perform MD simulations at a sliding velocity of $10^{-6}$. The running time was less than a week. However, without the aid of hyperdynamics, the simulations at lower sliding velocities ($<10^{-6}$) are not attainable because whenever we lower the sliding velocity by a factor of 10, we need to increase the running time by the same factor. Using a bias potential constructed using the eigenvalue and the local slope, the simulations at a sliding velocity of $10^{-7}$ were attainable.

All the data measured from the various methods show close agreement with the trend line obtained from the continuous method on a single processor within the standard deviation shown as the error bars in figure 13. Moreover, as expected from the modified Tomlinson model [23], the lateral force exhibits the logarithmic dependence on the sliding velocity within the range of the parameters used in this simulation study.

## 4. Conclusion

We have devised a novel scheme to accelerate the MD simulations for driven systems extending the original hyperdynamics method. Combined with a parallel algorithm simultaneously running systems at different slider positions on multiple processors, this extended hyperdynamics method has been applied to the frictional sliding of the 2-dimensional and 3-dimensional AFM models.

The validity of the methodologies was well verified by comparison with conventional MD simulations. First, we observed that the stepped sliding serves as a reasonable approximation for continuous sliding, and the simulation results using the parallel methodology and hyperdynamics showed close agreements with the simulation results of the conventional method. Moreover, both 2-D and 3-D simulations showed that the average of the lateral forces at the transitions have the logarithmic dependence on the sliding velocity as predicted from the modified Tomlinson model.

The sliding velocities used in experiment (nm/s ~ μm/s) and MD simulation (~ m/s) are different by several orders of magnitude and this difference cannot be completely overcome purely by parallelization methods such as the parallel replica method. The method applied here gains acceleration both from the boost factor in the hyperdynamics and from the parallel algorithm. Therefore, with this combined method we anticipate that it will be possible to simulate real systems investigated in AFM experiments over a comparable range of sliding velocities.

**Acknowledgements**

WKK and MLF acknowledge support of the NSF program on Materials and Surface Engineering under Grants CMMI-0510163 and CMMI-0926111 and the use of facilities at the University of Michigan Center for Advanced Computing and the Johns Hopkins University Homewood High Performance Compute Cluster.

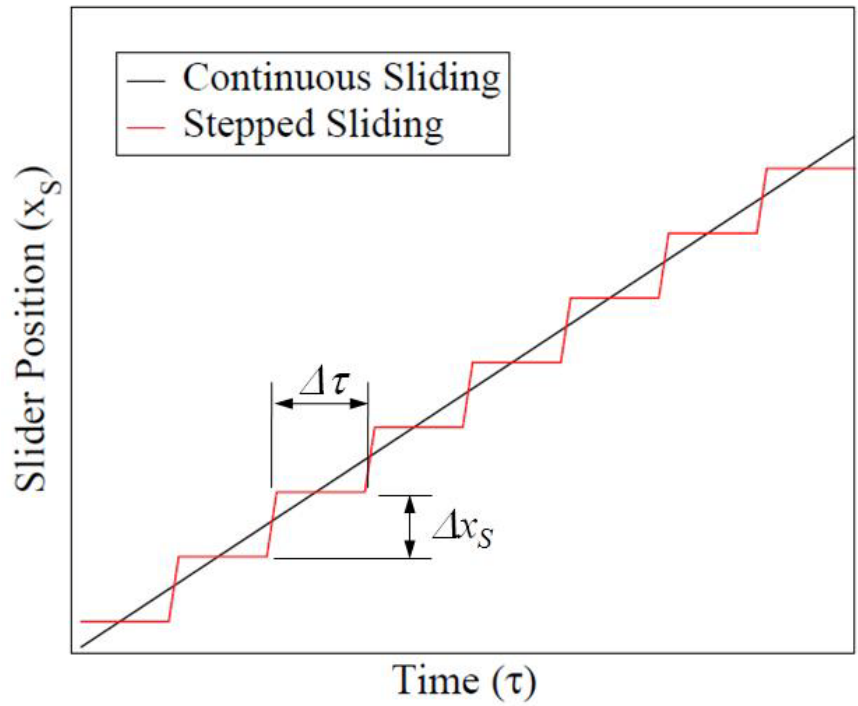

**Figure 1.** Slider position vs. time. The Black line represents the continuous change of the slider position in time. During stepped sliding (red lines) the slider position is fixed for $\Delta\tau$ and updated by $\Delta x_S$ after this time period has elapsed.

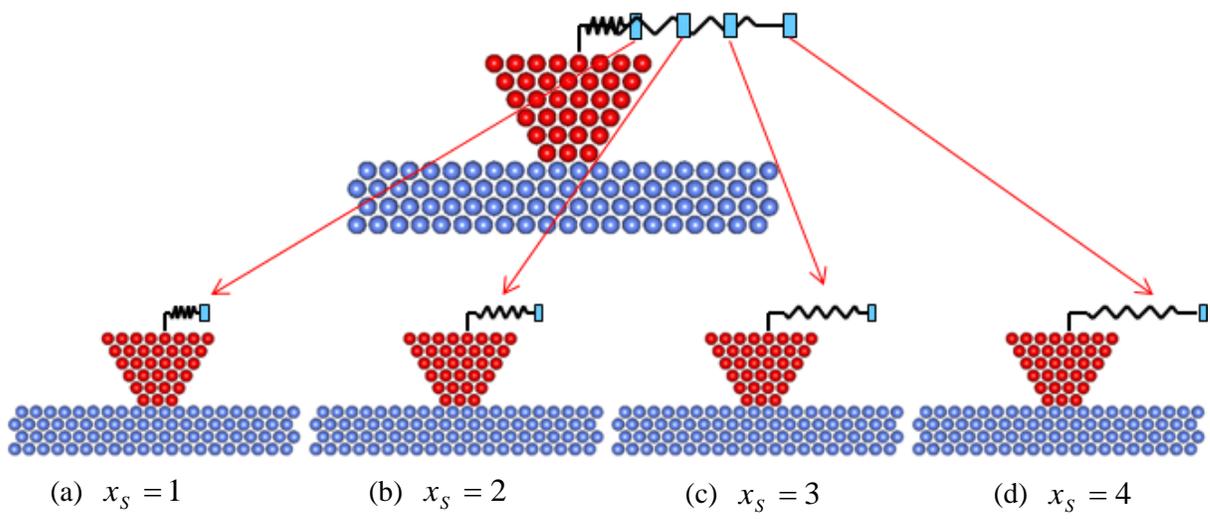

**Figure 2.** An illustration of the parallel distribution method.

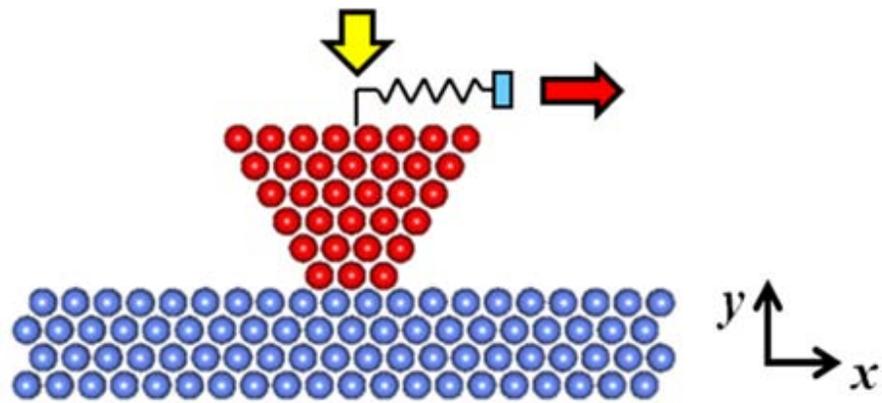

**Figure 3.** A diagram of a 2-dimensional AFM model consisting of a tip and a substrate. The tip atoms are shown in red and the substrate atoms are shown in blue. The top layer of the tip is pulled by a spring, which is attached to a slider moving in the positive x direction (the red arrow), and pushed by a normal force expressed as the yellow arrow.

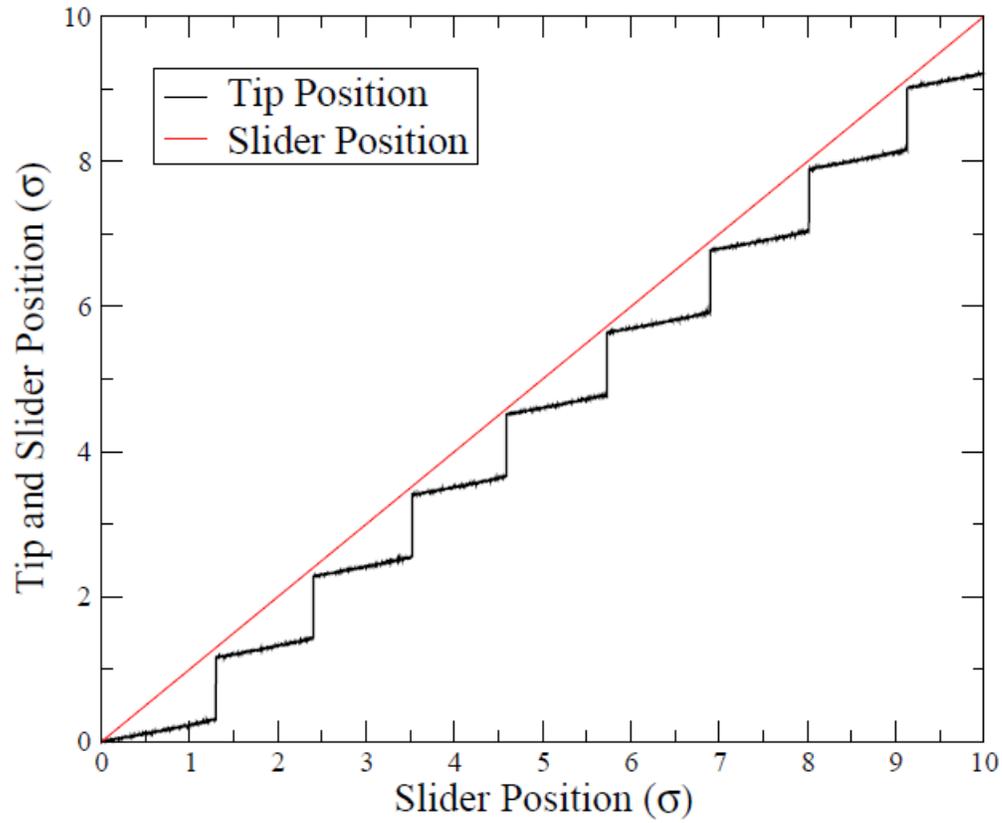

**Figure 4.** Tip position as a function of slider position calculated from the 2-D model simulated at $v_S = 10^{-6}$ and $T = 0.01$. Slider position is also plotted for comparison.

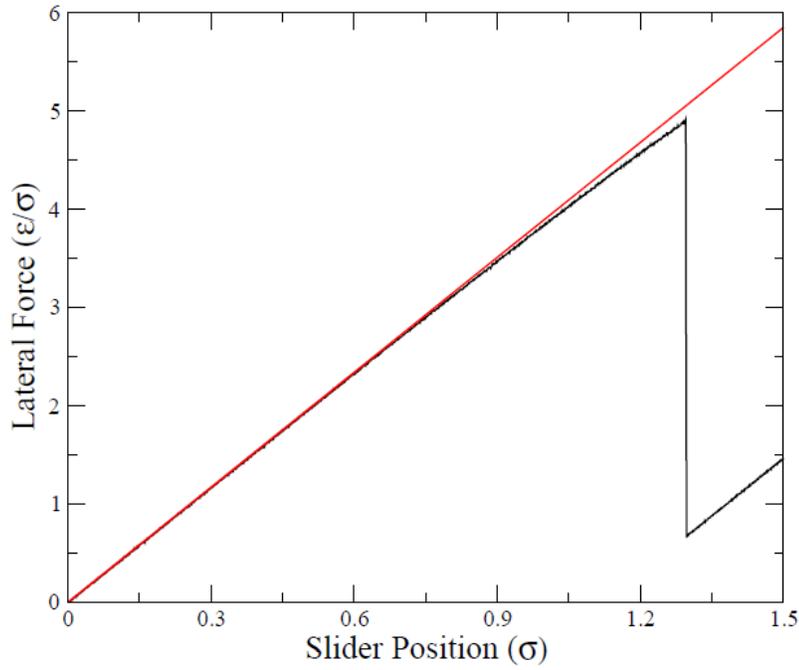

(a)

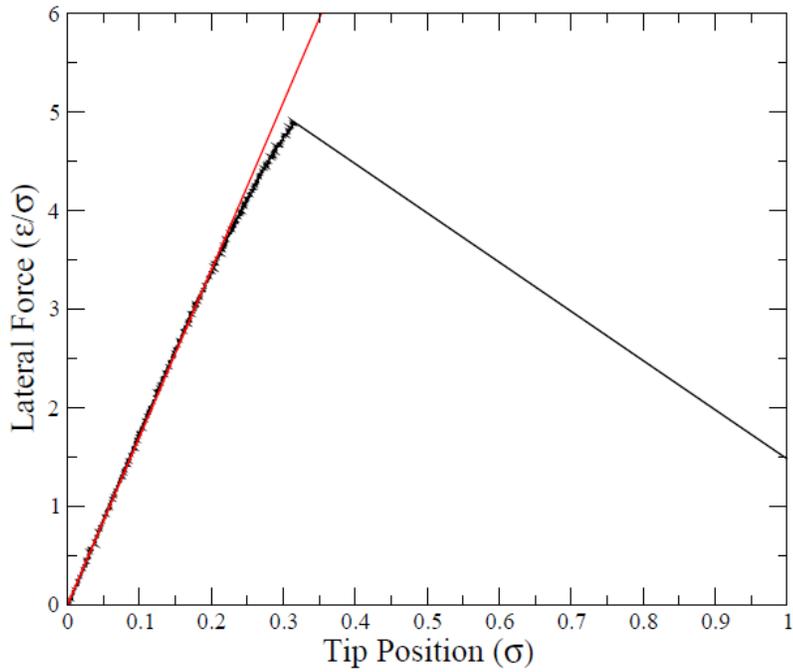

(b)

**Figure 5**. Lateral forces (black curves) calculated from the 2-D model simulated at $v_S = 10^{-6}$ and $T = 0.01$ and linear fittings (red straight line) extended from initially linear portion (a) Lateral force vs. slider position and (b) Lateral force vs. tip position.

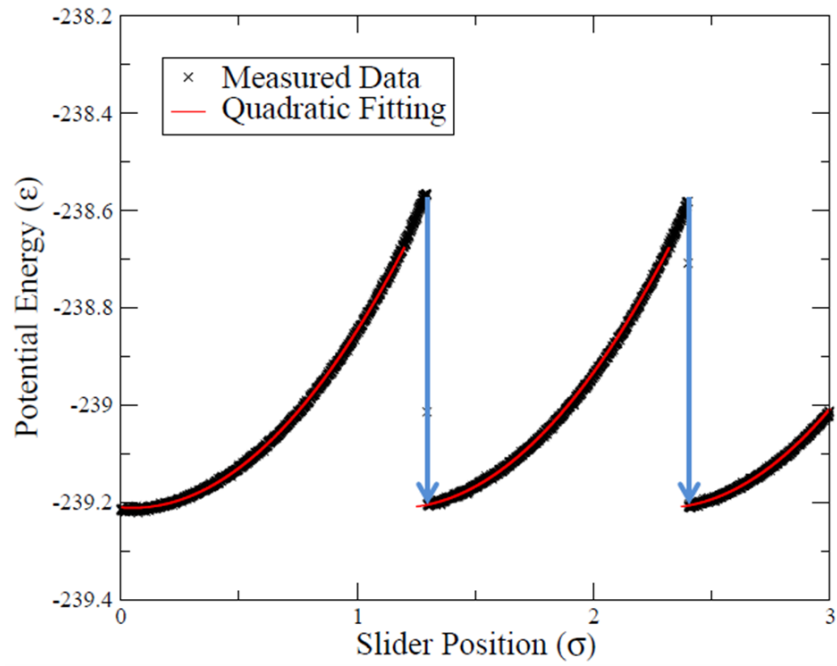

(a)

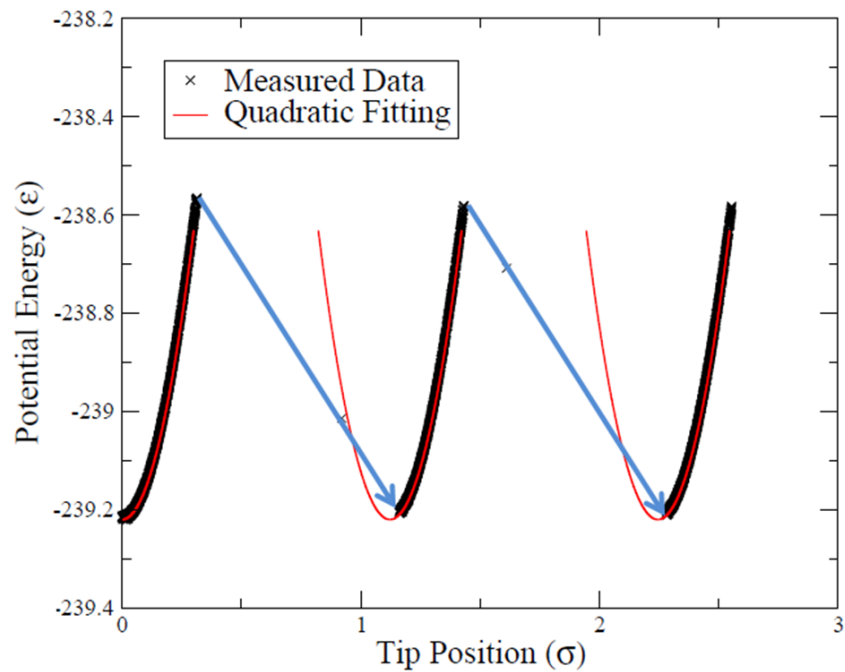

(b)

**Figure 6.** Potential energy curves (black curves) obtained from the 2-D model simulated at $v_S = 10^{-6}$ and $T = 0.01$, and quadratic fittings (red curves). The discontinuous points are connected by blue arrows. (a) Potential energy vs. slider position and (b) Potential energy vs. tip position. The blue arrows indicate transitions.

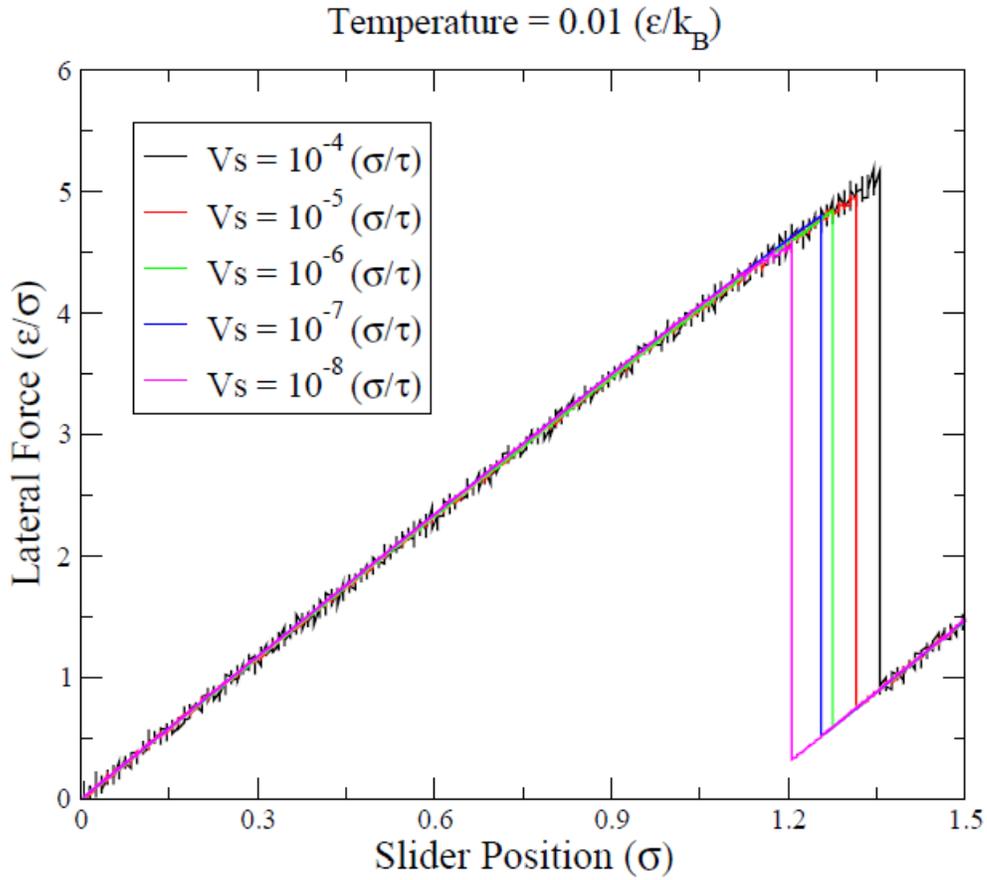

**Figure 7**. Sliding velocity dependence of lateral force obtained from the 2-D model simulated at $T = 0.01$. Lateral forces are shown as functions of slider position at five different sliding velocities ($v_S = 10^{-4}, 10^{-5}, 10^{-6}, 10^{-7}, 10^{-8}$)

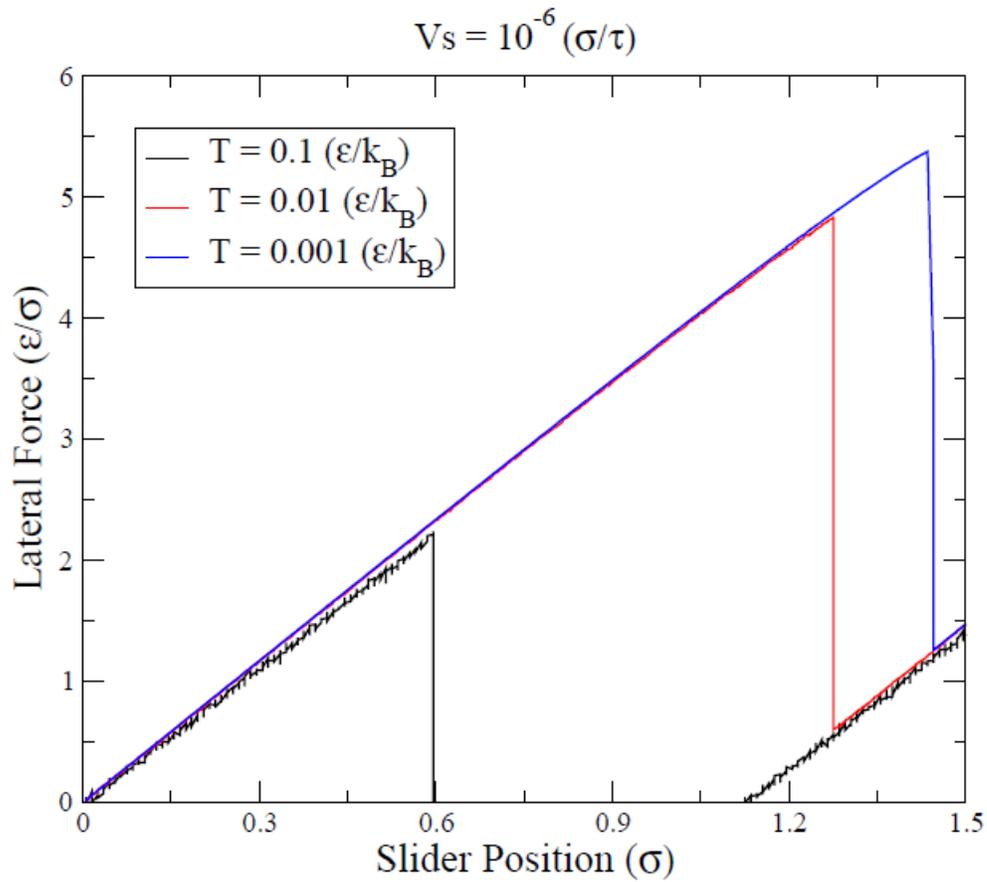

**Figure 8.** Temperature dependence of lateral force obtained from the 2-D model simulated at $v_S = 10^{-6}$. Lateral forces are shown as functions of slider position at three temperatures ($T = 0.001, 0.01, 0.1$).

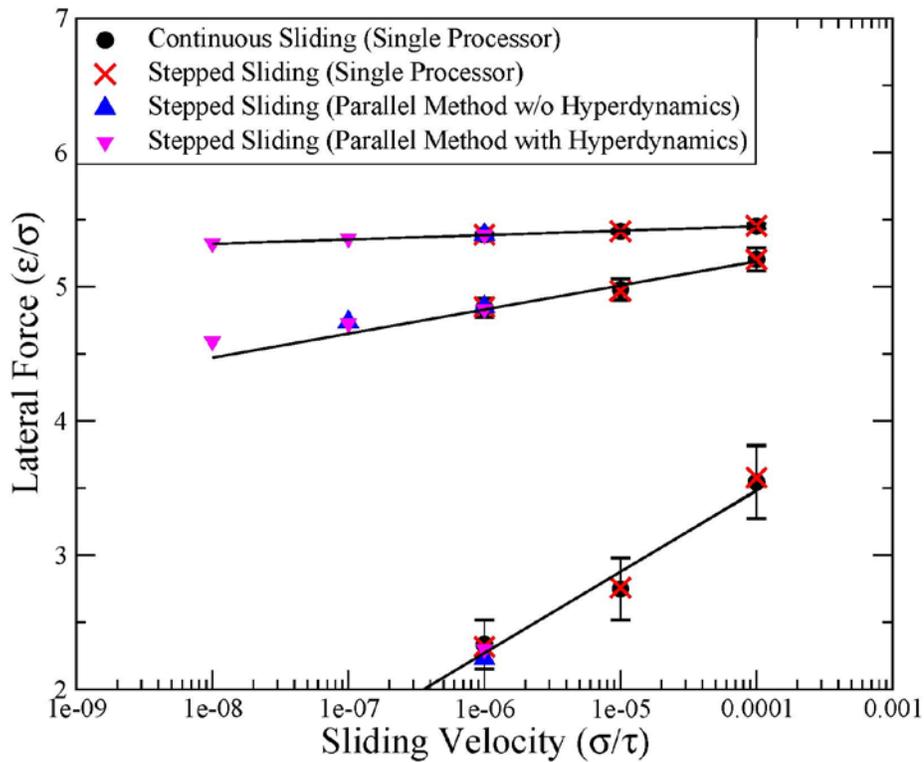

**Figure 9.** Lateral forces as functions of sliding velocity obtained from the 2-D model simulated at three different temperatures ($T = 0.001, 0.01, 0.1$) using four different methods. The straight trend lines and the error bars (the standard deviation) are obtained from the data of the continuous sliding simulations.

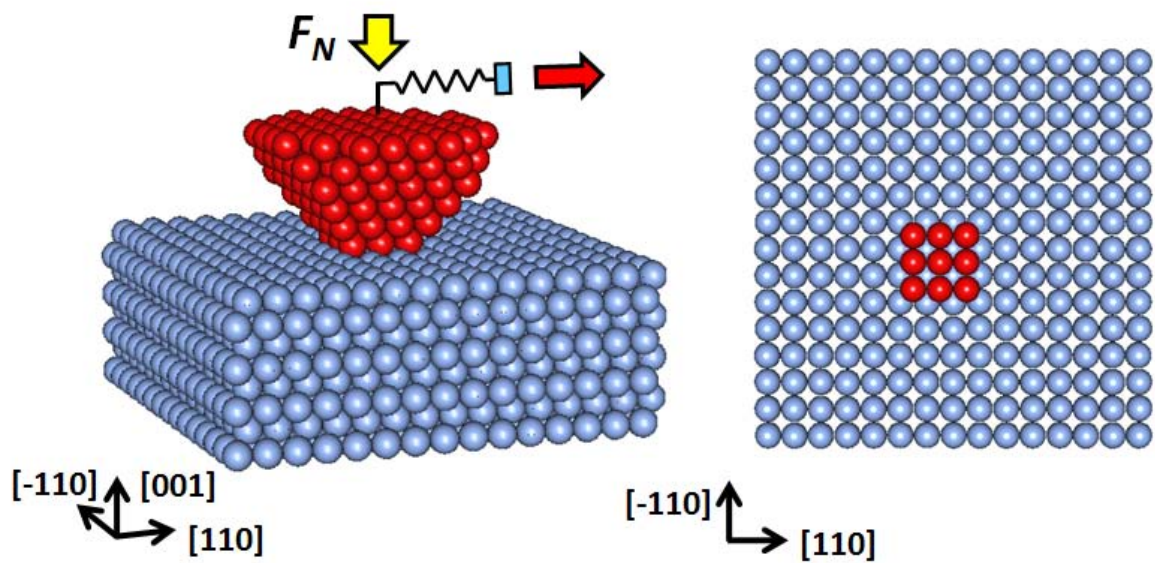

**Figure 10**. A diagram of a 3-dimensional AFM model consisting of a tip and a substrate. The atoms in the tip are shown in red, and the atoms in the substrate are shown in blue. The top layer of the tip is pulled by a spring moving in the sliding direction (the red arrow), and pushed by a normal force expressed as the yellow arrow.

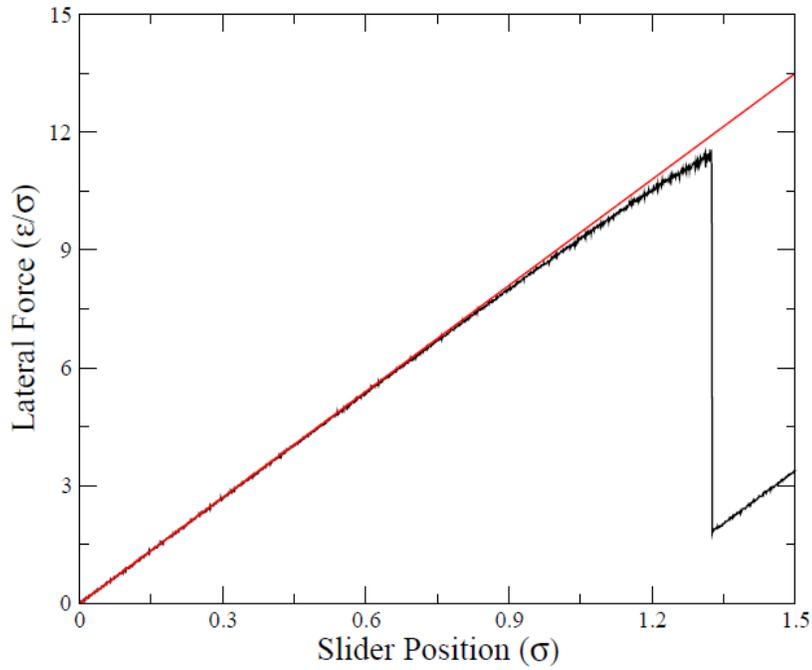

(a)

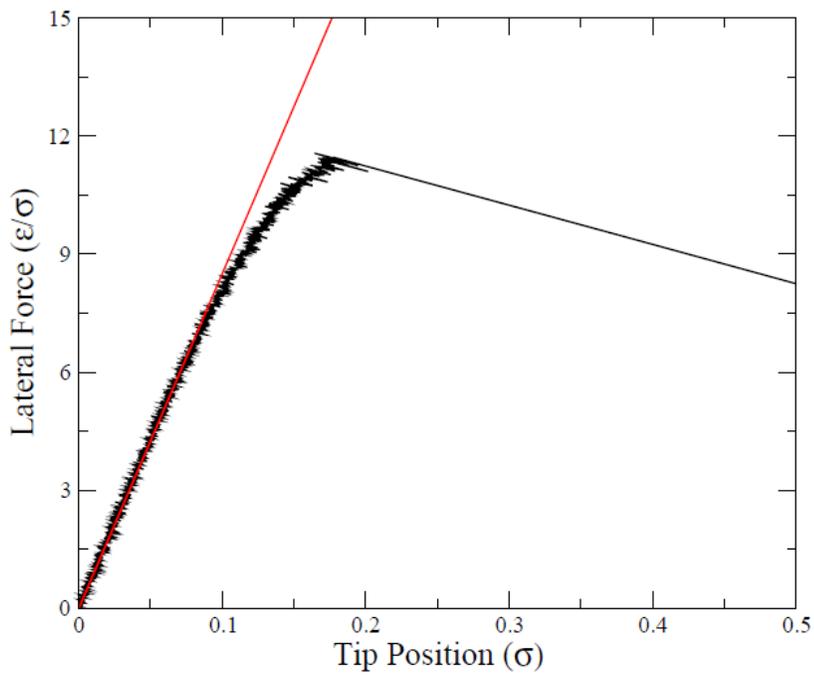

(b)

**Figure 11**. Lateral forces (black curves) calculated from the 3-D model simulated at $v_S = 10^{-5}$ and $T = 0.01$ and linear fittings (red straight line) extended from initially linear portion (a) Lateral force vs. slider position and (b) Lateral force vs. tip position.

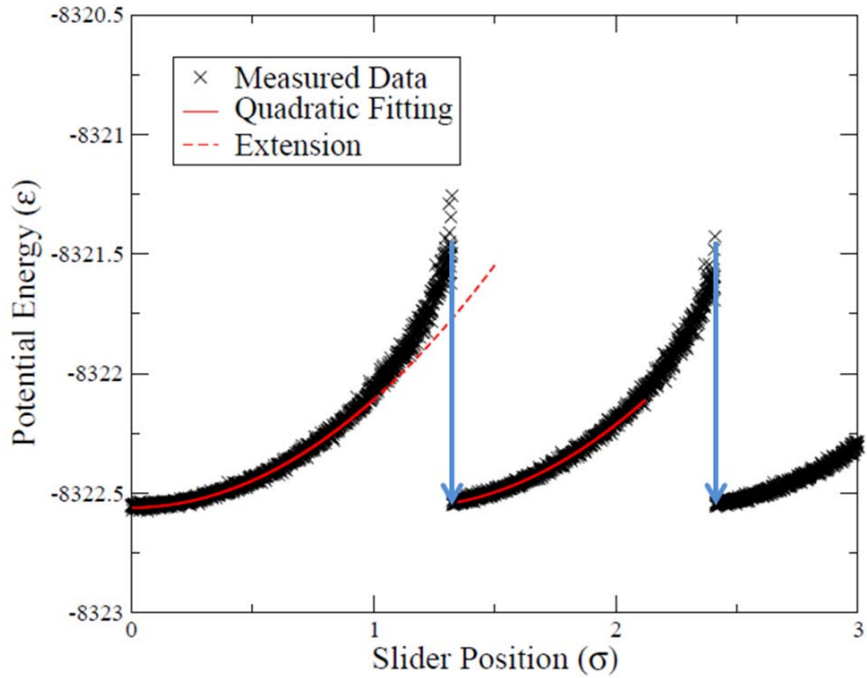

(a)

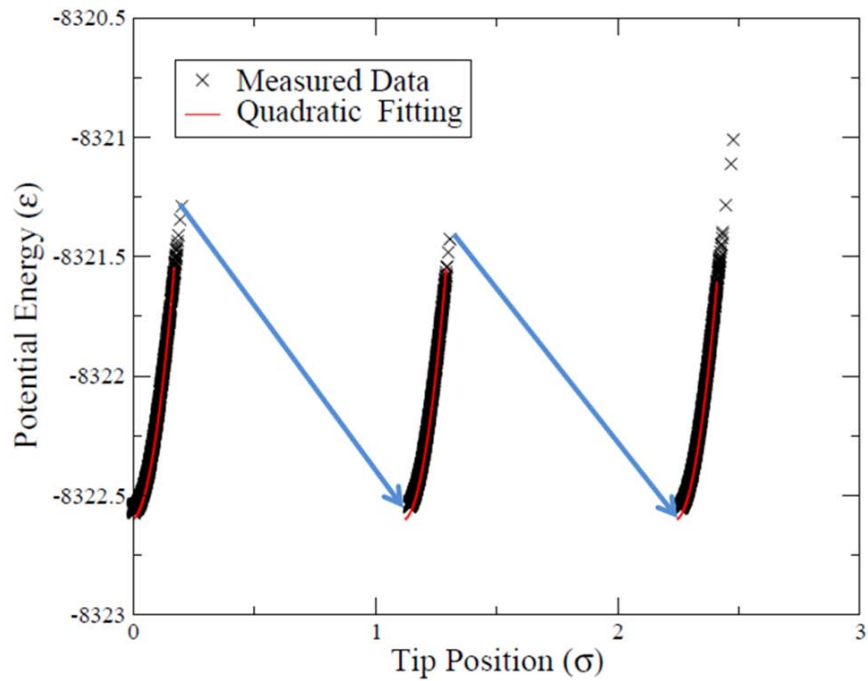

(b)

**Figure 12**. Potential energy curves (black curves) obtained from the 3-D model simulated at $v_S = 10^{-5}$ and $T = 0.01$, and quadratic fittings (red curves). The discontinuous points are connected by blue arrows. (a) Potential energy vs. slider position and (b) Potential energy vs. tip position. The blue arrows indicate transitions.

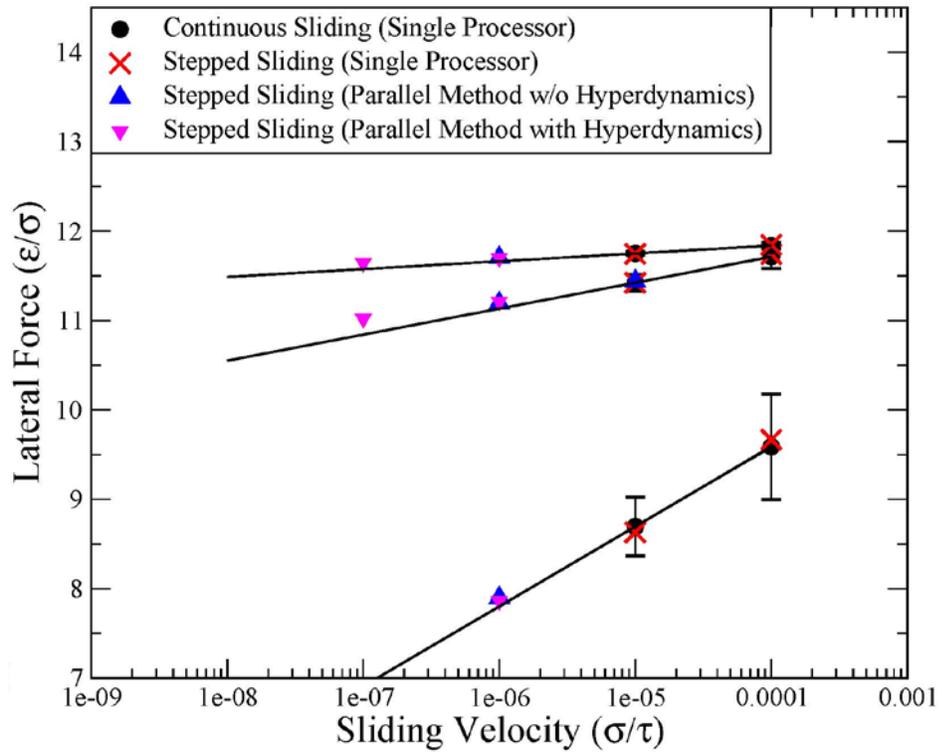

**Figure 13.** Lateral forces as functions of sliding velocity obtained from the 3-D model simulated at temperatures of 0.001, 0.01, and 0.1 using four different methods. The straight trend lines and the error bars (the standard deviation) are obtained from the data of the continuous sliding simulations.